%% file: MAIN.tex
\documentclass[conference]{IEEEtran}
\IEEEoverridecommandlockouts
\usepackage{xspace}
\usepackage{comment}
\usepackage{siunitx}
\usepackage{graphicx}
\usepackage{listings}
\usepackage{subcaption}
\usepackage{hyperref}
\usepackage{float}
\usepackage{url}
\usepackage[dvipsnames]{xcolor}

\begin{document}

\definecolor{mGreen}{rgb}{0,0.6,0}
\definecolor{mGray}{rgb}{0.5,0.5,0.5}
\definecolor{mPurple}{rgb}{0.58,0,0.82}
\definecolor{backgroundColour}{rgb}{0.95,0.95,0.92}

\lstdefinestyle{PythonStyle}{
    language=Python,
    backgroundcolor=\color{backgroundColour},
    commentstyle=\color{mGreen},
    keywordstyle=\color{magenta},
    numberstyle=\tiny\color{mGray},
    stringstyle=\color{mPurple},
    basicstyle=\ttfamily\footnotesize,
    breakatwhitespace=false,
    breaklines=true,
    captionpos=b,
    keepspaces=true,
    numbers=left,
    numbersep=5pt,
    showspaces=false,
    showstringspaces=false,
    showtabs=false,
    tabsize=2
}
\lstset{style=PythonStyle}

\newif\ifdraft
\draftfalse
\ifdraft
  \newcommand{\ian}[1]{{\textcolor{red}{ Ian: #1 }}}
  \newcommand{\logan}[1]{{\textcolor{blue}{ Logan: #1 }}}
  \newcommand{\ryan}[1]{{\textcolor{magenta}{ Ryan: #1 }}}
  \newcommand{\greg}[1]{{\textcolor{purple}{ Greg: #1 }}}
  \newcommand{\rajeev}[1]{{\textcolor{orange}{ Rajeev: #1 }}}
  \newcommand{\ganesh}[1]{{\textcolor{green}{ Ganesh: #1 }}}
  \newcommand{\kyle}[1]{{\textcolor{teal}{ Kyle: #1 }}}
  \newcommand{\yadu}[1]{{\textcolor{brown}{ Yadu: #1 }}}
  \newcommand{\frank}[1]{{\textcolor{brown}{ Frank: #1 }}}
\else
  \newcommand{\ian}[1]{}
  \newcommand{\logan}[1]{}
  \newcommand{\ryan}[1]{}
  \newcommand{\greg}[1]{}
  \newcommand{\rajeev}[1]{}
  \newcommand{\ganesh}[1]{}
  \newcommand{\kyle}[1]{}
  \newcommand{\yadu}[1]{}
  \newcommand{\frank}[1]{}  
\fi

\newif\ifrevision
\revisionfalse
\ifrevision
  \newcommand{\revise}[1]{{\color{red} #1}}
\else
  \newcommand{\revise}[1]{#1}
\fi

\newfloat{code}{htbp}{lop}
\floatname{code}{Listing}

\pdfstringdefDisableCommands{%
  \def\\{}%
}
  

\title{Colmena: Scalable Machine-Learning-Based Steering of Ensemble Simulations for High Performance Computing}

\pagestyle{plain}

\input authors
\maketitle


\begin{abstract}
Scientific applications that involve simulation ensembles can be accelerated greatly by using experiment design methods to select the best simulations to perform. Methods that use machine learning (ML) to create proxy models of simulations show particular promise for guiding ensembles but are challenging to deploy because of the need to coordinate dynamic mixes of simulation and learning tasks. We present Colmena, an open-source Python framework that allows users to steer campaigns by providing just the implementations of individual tasks plus the logic used to choose which tasks to execute when. Colmena handles task dispatch, results collation, ML model invocation, and ML model (re)training, using Parsl to execute tasks on HPC systems. We describe the design of Colmena and illustrate its capabilities by applying it to electrolyte design, where it both scales to \num{65536} CPUs and accelerates the discovery rate for high-performance molecules by a factor of 100 over unguided searches.
\end{abstract} 


\section{Introduction}

\revise{High performance computing (HPC) campaigns involving repeated runs of simulations codes are being applied widely in science and engineering, for example to simulate molecules or proteins in different configurations to build models of how properties change with temperatures (model fitting)~\cite{wu2016clusterexpansion} or to evaluate many engine designs to find those with optimal efficiency (optimization)~\cite{probst2019superlearner}.
While experiments themselves are independent and can be run in parallel, 
fixed computing budgets and ever larger search spaces make it increasingly important to use results from completed experiments to steer such campaigns, i.e., to inform decisions about which experiments to perform next.
Machine learning (ML) is an emerging tool for writing steering applications that can learn from new data faster than can human experts.}


The core of algorithms for steering ensemble simulations, in broad terms, is a process that selects inputs to simulations and determines allocations of resources to tasks.
Steering policies vary significantly in complexity. 
For example, 
genetic algorithms select new experiments by combining characteristics of previous simulations, typically dedicating resources equally among 
experiments~\cite{wozniak2018supervisor}, while
Optimal Experimental Design (OED) methods determine new experiments based on the predictions and uncertainty of an internal statistical model, with resources allocated to tasks using methods such as all-to-one task, batches of experiments~\cite{shali2018batchactive}, and streaming~\cite{kandasamy2018thompson}. 
Some methods deploy only a single type of task, and others can select among different task types or levels of accuracy~\cite{peherstorfer2018multifidelity}.

Steering algorithms that employ ML are particularly problematic to deploy on HPC due to the need to manage the execution of not only simulations but also steering (e.g., model update, experimental design) tasks.
Achieving good results requires balancing two often conflicting needs: performing steering tasks often enough, and fast enough, to enable good choices of simulation tasks, and maintaining high HPC utilization.  
At a simple level, both challenges can be solved by re-allocating resources---a common capability of many workflow systems.
On a deeper level, the challenges involve more subtle questions, such as
how many resources to allocate to steering and when to retrain ML model(s).
For example, one may want to enforce a limited budget for steering tasks or decide when to retrain models by comparing simulation outputs to ML predictions.
The problem of expressing such policies and deploying them to HPC requires much innovation.

Existing methods for steering HPC simulations are purpose-built tools that combine a specific class of steering algorithm with methods for deploying computations across resources.
An early example, Nimrod/O~\cite{abramson2000nimrod}, used  built-in optimization algorithms to choose which simulations to schedule over distributed clusters.
More recently, the CANDLE Supervisor library~\cite{wozniak2018supervisor} uses the Swift/T workflow engine~\cite{wozniak13swiftt} to distribute tasks selected by an optimization algorithm that determines new tasks after results are placed in an output queue~\cite{wozniak2018supervisor}.
Rocketsled works by adding a special ``steering'' task at the end of a Fireworks workflow definition that submits new work after completion (e.g., to launch a Bayesian optimization task)~\cite{dunn2019rocketsled}.
Variations of these patterns exist in other steering tools, including LibEnsemble~\cite{libEnsemble}, CAMD~\cite{montoya2020camd}, DeepHyper~\cite{balaprakash2018deephyper}, SuperLearner~\cite{probst2019superlearner}, and DeepDriveMD~\cite{lee2019deepdrivemd}.
Each steering system uses a different way to express planning policies and are integrated differently with workflow engine, all of which provide informative initial steps towards learning how to best steer ensembles with AI.
Further advancement in ensemble simulations will require systems that provide greater flexibility in policies for orchestrating machine learning and simulation tasks together.

We present Colmena, a general-purpose library for steering ensembles of experiments on HPC computing systems.
Colmena is an open-source Python code that permits writing complex agents for steering ensembles of simulations and executing them across diverse computational resources, with a particular focus on applications that use ML to design computational campaigns.
In this paper, we formalize the experiment steering process and describe the design principles for the Colmena library.
We then demonstrate the ability to use Colmena to steer simulations on \num{1024} nodes (\num{65536} cores) and illustrate its use on a molecular design challenge. \greg{We go up to \num{2048} in one experiment. Are we saying \num{1024} here because the \num{2048} was just for inference? We also mention "up to \num{1024}" in conclusion.}
We intend that Colmena will facilitate experimentation with advanced algorithms for steering experiments across diverse computing resources and, eventually, automated laboratories.

\revise{
The main contributions of this paper are:

\begin{itemize}
    \item An abstract formulation of the computational campaign steering problem.
    \item The introduction of Colmena, a Python library for steering computational campaigns on HPC.
    \item A demonstration of using Colmena to design molecular materials using quantum chemistry and ML.
    \item An analysis of the scaling performance of Colmena on a Cray~XC40.
\end{itemize}
}

\section{The Problem}
We first provide an abstract definition of the problem that we seek to solve and then describe the example application used in our experiments.

\subsection{Abstract Formulation}\label{sec:formulation}


Let us assume that we have a (typically large) set of \textbf{entities}, $e\in \mathcal{E}$, each with \textbf{properties} $p \in \mathcal{P}$; a set of \textbf{assays} (e.g., simulations or laboratory experiments) $a\in \mathcal{A}$, each of which can be used to estimate a property $P(a) \in \mathcal{P}$ of an entity; and a \textbf{scoring function} $\mathcal{S}$, that when applied to available data on an entity's properties returns either a numeric score or $\phi$ if data are inadequate to assign a score. Note that multiple assays may exist for the same property, each with different cost and accuracy characteristics.

Given $\mathcal{E}$, $\mathcal{P}$, and $\mathcal{A}$, we can determine entity properties by performing a series of \textbf{tasks}, each involving the application of an assay $a\in \mathcal{A}$ to an entity $e\in \mathcal{E}$ to obtain an estimated value $v$ for property $p=P(a)$. 
Given a record $D$ of such tests, with each $d\in D$ defined by a tuple $(e, a, p, v)$, we can assign a score to $D$. For example, we might define the score as that of
the single highest-scoring entity: 
\[
V(D) = \max_{x\in \mathcal{E}} S( \{d : d\in D\ \text{and}\ d.e=x\} )
\]
We can also determine the cost incurred to produce $D$ by summing the costs incurred to obtain each value
\[
C(D) = \sum_{d \in D} c(d)
\]



\textbf{Experiment design problem.}
When the number of entities, $|\mathcal{E}|$, is large and/or assays are expensive, 
it becomes impractical to evaluate every possible entity-property combination.
The quality of the answers obtained then depends on which assays have been performed. Thus we have an experiment design problem. If $\mathcal{D}$ is every possible combination of tests, and $B$ is a resource bound, then we want to identify a set of tests $D$ such that:
\[
\max_{D\in \mathcal{D}}  V(D) : C(D) \le B
\]

The order in which tests are performed then matters.
For example, do we focus on lower-cost assays that may identify promising entities, or on higher-cost assays that may confirm (or eliminate) promising entities? 



\textbf{Static vs. learned assays.}
We  distinguish between \textit{static assays}, which have fixed behavior over the course of an experiment (e.g., a simulation code) and \textit{learned assays}, which can be improved as more data are added to the record. An example of the latter is an ML model that approximates results obtained from an expensive simulation.

\textbf{Training.}
This additional kind of task is used to generate a new version of a learned assay given the current record: $a' = \texttt{retrain}(a, D)$. 
As ML model accuracy generally increases with training data quantity and diversity, we have another dimension to our experiment design problem---whether to perform assays designed to increase training data diversity or to characterize promising entities.

\textbf{Generating candidates}.
If the number of entities is large or even innumerable (e.g., all possible polymers),
we may introduce a generator $G$ that when called produces one or more new candidate entities based on the record. 

\textbf{Decision problem}.
If actions are taken one at a time, then system state at each step is captured by the sets of known entities $E$, associated data $D$, and assays (including learned assays) $A$. 
Initially, each may be empty or alternatively may be prepopulated to provide some initial knowledge of the problem.
At each step, the next action 
is one of: 
\begin{itemize} 
    \item 
    generate one or more new entities, $e=G(D)$;
    \item
    run a task $a(e)$ for some $a$ 
    and $e$; or
    \item
    (re)train a learned assay $a$ to generate a new $a'$. 
\end{itemize}

\subsection{Our Example Application}\label{sec:example-intro}
As an illustrative example, we apply Colmena to a problem in electrolyte design for next-generation batteries.
In this application, entities are molecules; properties of interest include atomization energy, ionization potential, toxicity, stability, and synthesizability; assays include a variety of computational methods with varying costs and accuracies;
and a scoring function might impose toxicity and stability thresholds and then sum the other properties.

More specifically, we present results for a version of this problem that involves a fixed search space of molecules and a single property to optimize: specifically, 10\textsuperscript{5} molecules (represented as SMILES strings) from the QM9 dataset
\cite{QM9-article,QM9-web};
a single property, ionization potential;
and two assays, namely a quantum chemistry (QC) simulation and an ML model; and
ionization potential as the quantity to maximize.
While simple, this configuration allows us to explore many relevant tradeoffs. 


We implement these two assays as follows. For the QC assay, we use the NWChem simulation code~\cite{valiev2010nwchem}.  We first parse the SMILES string and generate an approximate geometry using RDKit~\cite{landrum2006rdkit}. We then use NWChem through the QCEngine Python interface~\cite{smith2020qcarchive} to compute the equilibrium geometry for the neutral and oxidized molecule and then compute the vibrational modes for each molecule. All computations are performed at the B3LYP/3-21G level of accuracy and typically require six node-hours per molecule on four nodes of the Theta system at the Argonne Leadership Computing Facility, as described in \autoref{sec:theta}.

The ML assay uses an ensemble of message-passing neural networks (MPNNs)~\cite{gilmer2017neural} implemented in Tensorflow, each trained using a different subset of the training data. 
We employ an ensemble of models to produce both a mean and an estimate of model uncertainty for each prediction. The initial ensemble of 16 MPNN models were trained using \num{2563} oxidation potentials computed with the QC assay.
We use this additional dataset and any new data in subsequent retraining tasks, which we limit to 15 minutes on a single node. To apply the ML model, we first use RDKit to parse the SMILES string, featurize the data in a form used by a Tensorflow ML model, and then evaluate the MPNN ensemble. It requires $3 \times 10^{-6}$ node-hours to evaluate a single molecule, which equates to 100 molecules per node-second.

\section{Our Approach}


We formulated the decision problem abstractly as a sequential process, where planning and simulation tasks are performed serially.
However, in order to use highly parallel computers to accelerate the exploration process, we want to allow simultaneous execution both of \textit{multiple instances of the same activity} (for example, applying an assay to multiple entities) and of \textit{different activities} (e.g., running a simulation-based assay, retraining an ML model with simulation results,
running an ML assay, deciding which entities to explore next).
Running multiple instances of the same activity at once is important because, 
at least in the applications that we consider here,
no single action can scale efficiently to use all of a large parallel computer.
Running different activities at the same time is important because
different actions vary greatly in their computational demands;
thus, strict sequencing would reduce parallel efficiency.
However, achieving high parallel efficiency is difficult in practice due to competing needs for efficient execution (demanding high parallelism, modest communication), 
resource management (dynamically reallocating resources),
and timeliness of information (making results from one computation available rapidly to computations deciding on next actions).
We have implemented Colmena to solve these challenges.

\subsection{Colmena Architecture}

\begin{figure}
    \centering
    \includegraphics[width=\columnwidth,trim=5mm 2mm 1mm 0mm, clip]{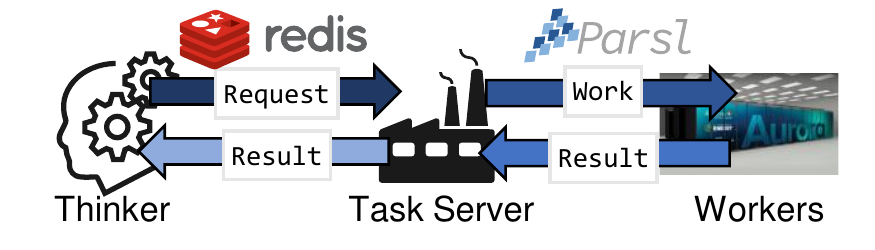}
    \caption{Illustration of the architecture of a Colmena application. The \textbf{Thinker} communicates task requests to the \textbf{Task Server}, which distributes computations across many \textbf{Workers}. Communication between Thinker and Task Server occurs via Redis; Task Server and Workers communicate via ZeroMQ channels managed by Parsl.}
    \label{fig:colmena}
\end{figure}

Applications built using Colmena are formed of three types of independent processes: a \textbf{Thinker}, a \textbf{Task Server}, and one or more \textbf{Workers} (as shown in Figure~\ref{fig:colmena}).

The user-supplied \textbf{Thinker} implements the decision-making policy used to generate new tasks, record assay results ($D$), and update ML models ($E$).
As described in \autoref{sec:formulation}, tasks can include performing a new assay on a specified entity, updating a learned assay, and generating new entities.
The Thinker communicates task requests to the Task Server; the results of those tasks, once available, are returned from the Task Server to the Thinker.
The Thinker makes decisions in response to results being communicated or other events (e.g., availability of compute resources).  


The \textbf{Task Server} matches each task request to the corresponding task definition (e.g., assay definition, ML retraining action) and dispatches the resulting task to an appropriate Worker.
The Task Server itself holds the assay definitions ($A$), information about available computational resources, and details of which assays can run on what resources.
Task requests are received from the Thinker and can be executed in any order.

Each \textbf{Worker} receives a sequence of tasks (\{\textit{task input}, \textit{task definition}\} pairs) from the Task Server. It executes each task that it receives and provides results back to the Task Server. 

From a decision perspective, the challenge is to 
allow these different activities to run with maximum concurrency and performance. 

\subsection{Colmena Implementation}
Our implementation of Colmena uses Redis~\cite{redis} for asynchronous communication between the Thinker and Task Server, and the Parsl~\cite{babuji19parsl} parallel programming
library to manage execution of tasks on the Workers.


\subsubsection{Thinker}
The Thinker decides which tasks are run and how to allocate resources among them.
It communicates with the Task Server by writing task requests to, or reading
results from, Redis queues. Requests and results are communicated as JSON
objects and contain the inputs to a task, the outputs of the task, and
a variety of profiling data (e.g., task runtime, time inputs received
by Task Server). We provide a Python API for the message format,
with utility operations such as accessing the
positional or keyword arguments for a task and serializing inputs and results.

A Thinker is composed of multiple Agents, implemented as threads.
As we demonstrate in \autoref{sec:exp}, separating the decision process into multiple cooperative agents simplifies expressing resource allocation and task selection strategies.
Each agent can communicate with the Task Server via Redis and with other agents via Python's threading library.
Each agent typically responds to a different kind of event (e.g., resources becoming available, completion of a certain type of assay).
For example, one agent might submits task requests to the Task Server 
from a task queue as resources become available,
and a second respond to results arriving from the Task Server by adding new tasks to the task queue.

\begin{code}
\begin{lstlisting}[language=Python,basicstyle=\ttfamily\small]
from colmena import thinker

PARALLEL_TASKS = 3
TOTAL_TASKS = 10

class Thinker(thinker.BaseThinker):
  def __init__(self, queues):
    super().__init__(queues)
    self.next_task = None
    self.results = []
        
  @thinker.agent
  def planner(self):
    # Submit initial tasks
    for _ in range(PARALLEL_TASKS):
      self.queues.send_task(random(),
                            task='simulate')
    # Until enough work is done
    while len(self.results) < TOTAL_TASKS:
      # Get ideas from the old results
      good_idea = f(self.results) 

      # Update the next task
      self.next_task = good_idea
        
  @thinker.result_processor
  def consumer(self, result):
    # Store the result in the database
    self.results.append((result.args,
                         result.value))
    # Submit the next task in queue
    self.queues.send_task(self.next_task,
                          task='simulate')

thinker = Thinker(queues)
thinker.run()
\end{lstlisting}
\caption{An example Thinker that implements the simple policy, ``run 10 tasks in total, three  at a time, generating a new task based on results obtained so far as each task completes.''
Its implementation comprises two agents, \texttt{planner} and \texttt{consumer}. The \texttt{planner} first sends three initial task requests to the Task Server, and then continually computes the best-possible next task to perform, given the current state. The second \texttt{consumer}, invoked whenever the Task Server sends notification that a task has completed, stores the result and submits the next task. The decorator \texttt{@agent} causes the \texttt{planner} to be launched as a thread when \texttt{thinker.run()} is called; the decorator \texttt{@results\_processor} indicates that \texttt{consumer} should be run, also as a thread, each time that a task completes.
}
\label{lst:thinker}
\end{code}

The \texttt{BaseThinker} class in Colmena simplifies writing Thinkers with multiple agents.
Agent processes run as separate Python threads that communicate with the Task Server 
via the Redis queue and with each other via Python threading tools.
Agents are implemented as methods of a \texttt{BaseThinker} subclass
and then marked with the \texttt{@agent} decorator, as illustrated in Listing~\ref{lst:thinker}.
There are also special-purpose types of agents, such as a ``result processor,'' that run after
certain conditions are met, such as a result completing.
The \texttt{BaseThinker} class provides a function that launches all of the decorated
functions to run concurrently.

Colmena also provides a class for monitoring, controlling, and allocating resources between different assays.
The resource allocation object, \texttt{ResourceTracker}, stores a fixed count of resources that are available to a Thinker class and how they are assigned into different pools.
Agent threads may query the availability of resources in each pool, 
acquire or release them, 
and change the levels of allocation between different pools.
The class is created using Python's Lock and Semaphore objects to ensure 
resource requests can occur and be fulfilled concurrently.


\subsubsection{Task Server}
The Task Server is a stateful entity that performs high-throughput task processing. It receives task requests from an \textit{inputs} queue and posts results asynchronously to an \textit{outputs} queue when a task is complete. The Task Server requires a robust and performant backend for managing
the execution of a diverse set of potential assays---from short-running, single-core
inference tasks to long-running, multi-node MPI simulations. 
Further, the Task Server should provide an intuitive way to represent
diverse assays, transparently serialize and transfer inputs/outputs
to/from Workers, dynamically provision 
resources in heterogeneous environments (e.g., clusters and clouds, CPUs and GPUs),
scale and make efficient use of large-scale systems, 
elastically adapt resource configurations in response to workload, 
and provide fault tolerance to reliably execute assays with 
performance monitoring, error capture, and checkpoint/retry. 

While there are several potential parallel and distributed computing toolkits and workflow systems (e.g., funcX~\cite{chard2020funcX}, Ray~\cite{mortiz2018ray}, Swift/T~\cite{wozniak13swiftt}) that could be used for this purpose, we implement the Task Server using Parsl~\cite{babuji19parsl}. 
Parsl is a parallel programming library for Python that extends Python's native \texttt{concurrent.futures} interface to enable high-performance, distributed computation and dataflow-based workflows.
Parsl provides the runtime infrastructure to execute Python tasks asynchronously
on various compute resources. It is able to serialize Python functions and
input arguments, transfer those functions and arguments to a remote system, 
execute the function in the configured Worker environment, and retrieve
results and errors.  Parsl's modular design and standard interfaces
enable workloads to be executed using various \textit{executors}, such
as via pilot jobs or a distributed MPI job, and by interacting
with various job schedulers and cloud APIs, such as Slurm, PBS,  
and Amazon Web Services (AWS). 
The ability for Parsl to interface with job schedulers and cloud APIs, in particular,
opens the possibility for writing application adjust the amount of resources
devoted to a problem during the course of an application (e.g., reducing the simulation resources while ML models are retraining).


Users create a Task Server by providing a list of tasks and specifying, using 
Parsl's Python-based notation, the target computational resources.
The tasks are defined as Python functions, which we wrap using Parsl's
\texttt{PythonApp} to allow the functions to be executed remotely.
Each assay can be mapped to different computational resources, making it possible 
to run assays on different resources (e.g., specialized hardware) or
use different types of Workers (e.g., single-node vs multi-node) on the same resource.

The Task Server itself is written as a multi-threaded Python process. 
An intake thread reads task requests from the inputs queue and
uses them to launch Parsl task executions.
Output threads write completed tasks to the result queue.



\subsubsection{Communication}
\label{sec:architecture-communication}

The Thinker and Task Server communicate via Redis queues, 
with distinct \textit{request}/\textit{result} queue pairs for different task types (e.g., different assays, ML training). 
The Thinker writes requests to the appropriate \textit{request} queue,
to be received by the Task Server; when task execution completes, the Task Server
writes the result to the corresponding \textit{result} queue, from which it is read by the Thinker.
This use of different queues for different task types
simplifies implementation of Thinkers with multiple sub-agents.

Upon receiving a \textit{task} request,
the Task Server creates and launches a corresponding Parsl task. Parsl uses a hierarchical communication model, with ZeroMQ channels to efficiently distribute tasks to its Workers. As the Task Server receives results from Workers over these
channels, it posts each to the appropriate \textit{result} queue.

For tasks with large input or result values, Colmena uses a \textbf{Value Server} to pass values directly from the Thinker to the Worker---bypassing the Task Server.
The Value Server uses Redis as the backend key-value store and exposes a lazy object proxy interface.
Lazy object proxies simplify interaction with the Value Server because (1) the proxies behave as the wrapped object so users do not need to modify code to accommodate the proxies, (2) the proxies automatically handle retrieving data from the value store once the data are first needed, and (3) the lazy aspect of the proxies can amortize communication costs with the Value Server.

Any arbitrary object $v$ can be wrapped in a proxy.
The process of wrapping $v$ involves placing $v$ into the Value Server and returning a proxy $p$ that stores the key associated with $v$ in the Value Server and some additional metadata.
Proxies behave like the underlying object, e.g., \texttt{isinstance(p, type(v))==True}.
The proxy $p$ is lazy in that it acts as a reference to $v$ until $p$ is accessed.
Thus, $p$ is cheap (in terms of serialization and communication costs) to include as a task input in the \{\textit{task input}, \textit{task definition}\} pair.
When first used, $p$ is \emph{resolved}, meaning $v$ is retrieved from the Value Server and stored inside $p$ such that $p$ can be used as $v$ would be.

Colmena will automatically proxy input and result values larger than a user-defined threshold, and/or users can manually proxy large objects.
The Value Server has a Worker-level cache to speed up tasks that reuse the same inputs (e.g., the model for ML inference tasks).
Proxies can be asynchronously resolved, allowing for the overlap of Value Server access and computation.
Colmena starts asynchronously resolving all proxies in a task's input prior to the task being executed on a Worker.
Thus, the communication with the Value Server is overlapped with the task's execution.
The start of a task often involves some initialization or importing of libraries, such that by the time a value is needed by the task, the corresponding proxy has already been resolved in the background.

\subsection{Measuring Communication Overheads}\label{sec:impl-lifecycle}

There are several places for communication overhead between a task selected
by a Thinker and the result being received by the Thinker again.
As illustrated in \autoref{fig:colmena}, each task request has two pairs of serialization/deserialization steps and four data transfer steps.
The time required for serialization and deserialization steps are recorded automatically and
transmitted along with the task data.
The communication times are also measured indirectly by the difference in time between when an object is written to and read from a Redis queue.
These times can be compared to the time spent executing the assay, which is also transmitted along with the task request and result objects.
The data are available both to aid scaling applications, and can also be used proactively by Thinker applications when planning tasks.

\section{Application Experiments}\label{sec:exp}

We conducted experiments to evaluate the performance of our molecular design application and then further studied the performance of components that were bottlenecks in the molecular design application.

\subsection{Application Description}


As introduced in \autoref{sec:example-intro}, our example application involves an ML-guided search of $10^5$ molecules for those that match one of our design criteria for molecular electrolytes, namely high ionization potential (IP).
At a high level, our application is an adaption of Bayesian Optimization to HPC.
We determine a score of each task using the Upper Confidence Bound (UCB), a value based on the mean and confidence interval of the predictions from an ensemble of MPNNs trained to predict the IP from the bonding network of a molecule.
UCB defines the highest scoring molecules as those with large means and large confidence intervals~\cite{lai1985ucb},
which are those likely to both have high performance and provide data that will improve the models (i.e., because they are in regions where the model is least certain).
Molecules are evaluated in order of descending scores.
The data from completed simulations are used to update the MPNNs and, through the updated models, produce better estimates of molecular properties.
Many simulations are performed in parallel as NWChem scales poorly for the small molecule sizes considered in this study.

More formally, the application uses two assays: a more expensive and accurate QC assay and 
an inexpensive but less accurate ML assay trained on QC results.
As we are using a pre-defined search space of molecules, no generator is needed to expand the set of molecules considered progressively during execution.
Instead, the application's Thinker maintains two data structures---a \textbf{molecule queue} of \{\textit{molecule}, \textit{ML-score}\/\} pairs, ordered by \textit{ML-score}, and a \textbf{results record} of \{\textit{molecule}, \textit{QC-score}\/\} pairs---that are 
manipulated by the following three pairs of agents (see \autoref{fig:molthinker}):

\begin{figure*}
    \centering
    \includegraphics[width=\textwidth,trim=0mm 0mm 2mm 0mm,clip]{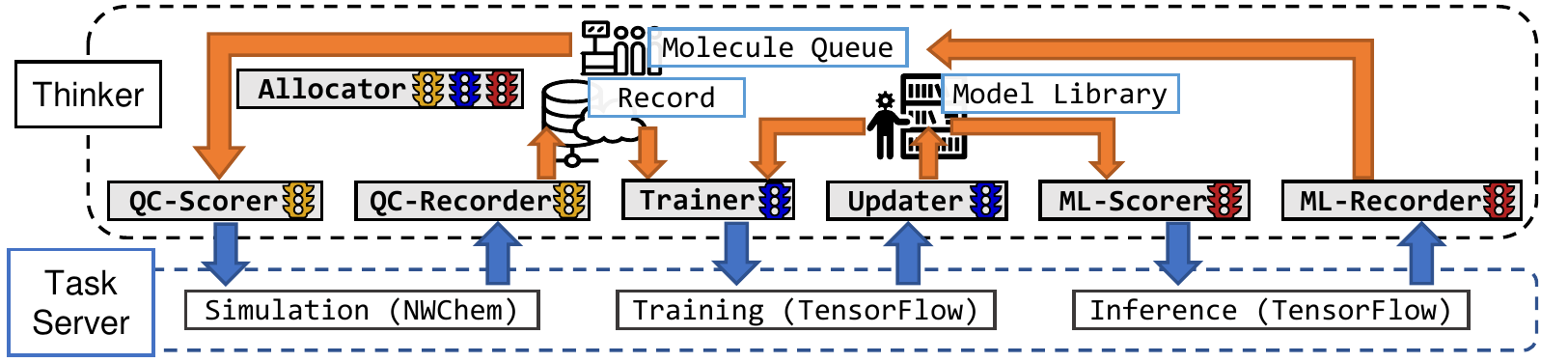}
    \caption{Implementation of molecular design application with Colmena. Agents within the Thinker application are gray boxes, and the three different available tasks are listed as white boxes. Blue arrows indicate communication between Thinker and Task Server; orange arrows illustrate how information flows between agents. Different colored traffic lights indicate resource pools used for different task types. The Allocator agent reallocates resources between different pools.}
    \label{fig:molthinker}
\end{figure*}

\begin{itemize}
\item 
The \textbf{Trainer} periodically sends to the Task Server a \texttt{retrain} task to retrain the ML model based on data in the results record;
the companion \textbf{Updater}, upon receiving results for such a task, updates the weights of the ML model.
\item 
The \textbf{ML-Scorer}, whenever the ML assay is updated, sends \texttt{ML-assay} tasks to the Task Server  to re-score every entity in the molecule list;
the companion \textbf{ML-Recorder}, upon receiving results of these tasks, computes the Upper Confidence Bound (UCB) for each and reorders the molecule list with the new information.
\item 
The \textbf{QC-Scorer} repeatedly removes a molecule from the front of the molecule list and sends a \texttt{QC-assay} task to the Task Server to determine its QC score;
the companion \textbf{QC-Recorder}, upon receiving results for these tasks, stores them in the results list if they pass validation.
\end{itemize}

The application's precise behavior thus depends (in addition to the total number of available resources) on the allocation of available resources to \texttt{retrain}, \texttt{ML-assay}, and \texttt{QC-assay} tasks: increasing the fraction allocated to \texttt{QC-assay} tasks leads to relatively more QC assays being performed, while increasing the fraction allocated to \texttt{retrain} and \texttt{QC-assay} tasks increases the timeliness of the ML-based scores, and thus in principle leads to more relevant QC assays being performed. 

The strategy for controlling the resource allocations is implemented as an \textbf{Allocator} agent.
The application's \textbf{Allocator} balances competing demands for resources with a policy that, after an initial set of ML
assay results are obtained with a pretrained ML model, 
(a) retrains the ML model each time that the results record increases in size by an amount $n\_retrain$, 
(b) reruns the ML assay on all entries in the molecule list whenever the ML model is updated, and 
(c) reallocates resources between \texttt{retrain}, \texttt{ML-assay}, and \texttt{QC-assay} tasks as needed to ensure that \texttt{retrain} and \texttt{ML-assay} tasks are run as fast as possible when generated, with resources otherwise being used for \texttt{QC-assay} tasks. 
Resource reallocations are performed by controlling the amount of requests sent to the Task Server, which generates requests to the Parsl backend to stop or start Workers.
Resources allocated to different purposes thus change in increments of the largest number of nodes needed for a single task, which in this case is four nodes for the \texttt{QC-assay} task.


In addition to the molecule list and results record already mentioned,
these processes also share a resource counter (used to track resource availability) and a library of ML models.


\subsection{Experimental Setup}\label{sec:theta}

We used the Theta supercomputer at the Argonne Leadership Computing Facility (ALCF)~\cite{harms2018theta},
a 11.69-petaflop system based on the second-generation Intel Xeon Phi ``Knights Landing'' (KNL) processor. Its 4392 nodes each have a 64-core processor with 16 GB MCDRAM and 192 GB of DDR4 RAM, for a total of \num{281088} cores; nodes are interconnected with a high speed Cray Aries network.

We performed 256-node and \num{1024}-node runs of the electrolyte design application on Theta. 
The application was configured so that the Thinker process ran on the machine-oriented miniserver (MOM) node, each NWChem task was allocated four nodes, and each Tensorflow task ran on one node each.
Tensorflow inference tasks were grouped into batches of 4096 for efficient multithreaded execution on each node.

\subsection{Application Evaluation}

We evaluate the performance of our molecular design application from the two perspectives of computational performance (specifically, computational efficiency) and the quality of the results obtained.

\subsubsection{Performance Evaluation}\label{sec:mdesign-scale}
We evaluate the performance of the application by measuring the fraction of the time worker processes spend performing the computational tasks requested by the thinker (e.g., simulation, ML inference) and not communicating work to/from the Task Server;
100\% indicates perfect performance.

\begin{figure}
    \centering
    \includegraphics[width=\columnwidth,trim=4mm 3.5mm 3.5mm 2.5mm,clip]{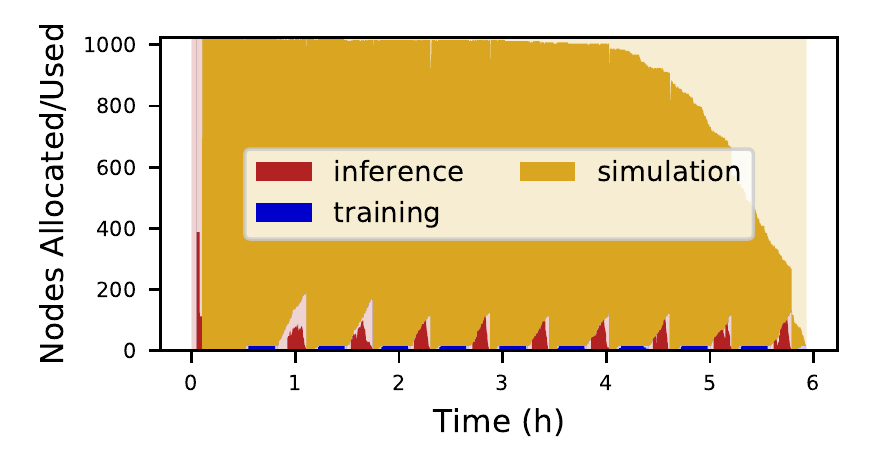}
    \caption{Allocation and utilization of compute resources during a run of the molecular design application on 1024 Theta nodes. Light shades indicate the nodes that are allocated to a problem; dark shades are nodes that are both allocated and are being used to execute assays. Different colors indicate different types of tasks. The large decrease in utilization at later times is a result of trailing tasks~\cite{armstrong2010scheduling}.}
    \label{fig:utilization}
\end{figure}

As shown in \autoref{fig:utilization}, we maintain near 100\% utilization for most of a \num{1024}-node run of our application. 
We note two major sources of under-utilization. 
The first is the start-up time 
for inference Workers, which is a median of 3 minutes.
The startup cost can be reduced by unpacking Python libraries to node-local memory before launching Parsl Workers~\cite{shaffer21lfm}.
The second source of under-utilization is simulation tasks that do not complete within the timescale of the job, which leads to the associated resources being counted as unutilized even though the calculations are running  (as verified via Parsl logs).
This source of under-utilization can be mitigated by periodically checkpointing simulation tasks or by splitting simulation tasks into smaller steps, such as computing the neutral geometry first and then computing the ionization potential.
The latter approach has the advantage that the Thinker can then use intermediate results to decide whether the computation should be continued (e.g., by deciding if it is likely to complete before the end of an allocation).

The overheads due to communication are minimal for \texttt{QC-Assay} tasks.
The median cost for launching a new simulation is 1015~ms: 620~ms for the result to be received by the Thinker, 35~ms for submitting the new task to the Task Server, and 360~ms for the task to be launched on a Worker.
The cost here is minimal (0.03\%) compared to the median simulation runtime of 3275~s.
We note that there is a significant variation between the largest cost, result communication, with that of failed tasks often 10$\times$ shorter than successful tasks.
The difference can be attributed to the amount of data transferred, with failed tasks typically sending 0.5~KB and successful tasks communicating a median of 4.3~MB.

The communication cost of the \texttt{ML-Assay} tasks is partly hidden by prefetching tasks to Workers, but we note two issues that lower utilization and could inhibit further scaling.
One is the startup time for the Workers used for ML assays.
We find that it takes a median of 175~s for a Worker to begin work after being sent its first task from the Task Server, due to the startup time noted earlier.
The second startup cost is the time it takes to compile a TensorFlow model before execution.
The first task run by a Worker requires a median of 100~s, and the second task requires only a median of 80~s. 
We observe similar startup issues for the training~tasks.

A second issue for inference is the time required to communicate results from a Worker to the Task Server.
The communication time for an inference task request is 500~ms (0.6\% of the median execution time), which can be hidden by prefetching work to the Workers. 
While this cost has not been a problem in the simulations run to date, it may become so at larger scales or problems.
As with the \texttt{QC-Assay} tasks, we associate the long communication times with transferring large data objects.
The training tasks require 30~s to communicate the $>50$~MB updated ML model, a small (3\%) fraction of the median run time of 850~s.

In summary, we find the Colmena Task Server fulfills the performance requirements required for our application.
We can change the allocation between node-parallel and single-node tasks during the run, with the latency dominated by the time to start a Python interpreter ($\sim$100~s) for single-node tasks.
The latency for launching new tasks after a simulation completes is low ($<$1s), though we note this latency can increase to several seconds for tasks with large inputs (e.g., entire deep learning models).
Overall, the system was able to maintain a total utilization of 85\% during the run with the largest source of underutilization resulting from the trailing tasks.

One approach to improve scaling in the future may be to have the Thinker application consider Worker start-up costs when allocating resources.  For example, it may be better to maintain a smaller number of nodes dedicated to the \texttt{ML-Assay} tasks, so as to leverage the benefits of ``warmed'' nodes for inference tasks, rather than repeatedly stopping and starting inference tasks over time.
A second approach is to optimize the communication of large results and requests, which we describe in the next sections.

\subsubsection{Molecular Design Performance}
We assess the performance of the application at solving our target problem by studying the scores of the molecules in the record over a run.
We perform several runs using 256 nodes where we either retrain the ML assay on-demand during the run (as in \autoref{fig:utilization}) or only once at the beginning of the run. 
For context, we also compare both versions of the application to a run where we select tasks randomly.

\begin{figure}[tb]
    \centering
    \includegraphics[width=\columnwidth,trim=4mm 4mm 3mm 3mm,clip]{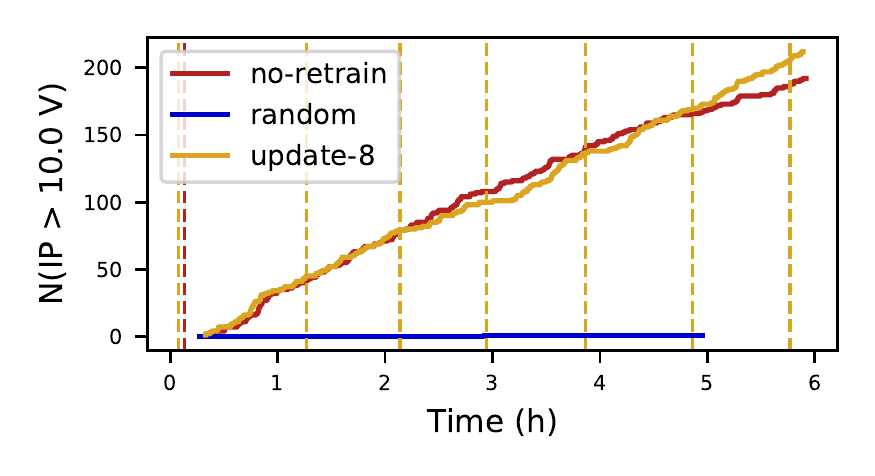}
    \caption{The number of high-performing molecules (i.e., with ionization potential greater than 10~V) evaluated over time by the Colmena molecular design application  when using three different Thinker strategies: 
    \textit{random}, which selects molecules for QC simulations at random;
    \textit{no-retrain}, which selects molecules for QC simulations based on an ML model that is trained once; and
    \textit{update-8}, which selects molecules based on an ML model that is retrained after 8 QC simulations complete successfully.
    Vertical dashed lines indicate the times at which the molecule list was reordered after model rerunning \texttt{ML-Assay} tasks. 
    }
    \label{fig:optimization-perf}
\end{figure}

\autoref{fig:optimization-perf} shows QC results as a function of time for Thinkers with different policies.
The two Thinkers that use ML models to select molecules for QC simulation perform much better than the one that selects molecules at random, finding over 100$\times$ more high-performing molecules with ionization potentials above 10~V.
Normalizing by the total number of molecules evaluated during the run, the random agent finds a high-performance molecule with a success rate of 0.5\%, whereas the success rates are 78\% and 64\% for the runs with and without retraining tasks.
In short, we find a significant advantage in using ML assays to prioritize the order in which we consider a molecular design space and also an advantage to reprioritizing the list of simulations during a run.
The application where we respond to new simulation results finds 10\% more molecules with large ionization potentials even though it spends 5\% less time on simulations.

The benefits of active learning are clearest at the end of the run, where the application uses data from the largest number of previous simulations to select new inputs.
In the last hour of each run, the average ionization potential of the application with retraining was significantly higher than the run without (10.5~V vs.\ 9.8~V), clearly illustrating that retraining leads to an improved ability to identify high-value simulations.
The effect can potentially be traced to improvements in the machine learning models.
The initial models have a mean absolute error (MAE) of 0.395~V on 185 molecules selected at random of the search space and the MAE of the models is reduced to 0.389~V by the second re-training event and 0.382~V by the last batch of the run.
The performance gains are subtle but the increased search performance illustrates how even minor improvements in a model can lead to an improved ability to select new molecules.
The challenge then becomes being able to perform model updates quickly.

Our application provides reasonable response times between when a simulation completes and its data are used to select the next simulations.
The time-to-solution for retraining is an average of 57~minutes between when a simulation completes and the task list is updated based on its results, in which time
an average of 63 new simulations are submitted---15\% of the 537 submitted during the entire run.
Such results demonstrate the need to dedicate specific resources to the ML tasks to keep up with the rate data can be produced by the simulation codes.
Dedicating fewer nodes to the retraining or inference task would slow how quickly the application responds to new data, to the detriment of the active learning process.

Our work demonstrates clear benefits for online training of ML models and opens up previously inaccessible opportunities for evaluating adaptive experimental design on HPC systems.
For example, our application presents an excellent use case for asynchronous, batched active learning strategies~\cite{kandasamy2018thompson}.
There is also significant room to devise optimal strategies for partitioning resources between simulation and learning tasks that are aware of the overhead for re-tasking nodes (e.g., start-up costs of machine learning workers).
We hope that Colmena and our example application provide a starting point for innovating new algorithms to steer ensemble simulations on HPC.

\subsection{Component Evaluation}
As discussed in \autoref{sec:mdesign-scale}, we observed that transferring large requests or result objects is a major source of communication costs, and the \texttt{ML-Assay} tasks are the most susceptible to the effect of this bottleneck.
Consequently, we first evaluate a system that optimizes large ($>10$~MB) result transfers and then evaluate its effect on \texttt{ML-Assay} tasks. This is the done through inclusion of the Value Server, which reduces the requirement of serialization and deserialization of task data. We  discuss the performance improvement through the inclusion of a Value Server using a synthetic problem. We then discuss the improvement that can be achieved in  real-world production runs for the electrolyte design problem. 

\subsubsection{Synthetic Application}\label{sec:synthetic-problem}
\label{sec:value-server-eval}



We built a synthetic application, SynApp, to permit the systematic evaluation of Colmena communication overheads.
This application uses a Thinker plus $N$ workers, one per node; the Thinker generates $T$ identical tasks, each with  duration $D$, unique (and thus non-cacheable) input of size $I$, and producing a result of size $O$. 
This Thinker first submits one task per worker and then continues to submit a new task each time that it receives a result, until $T$ tasks have been submitted.
We use this application to measure costs for different \{$T$, $D$, $I$, $O$, $N$\} combinations.

\begin{figure}
    \centering
    \includegraphics[width=\columnwidth,trim=2mm 2mm 2mm 2mm,clip]{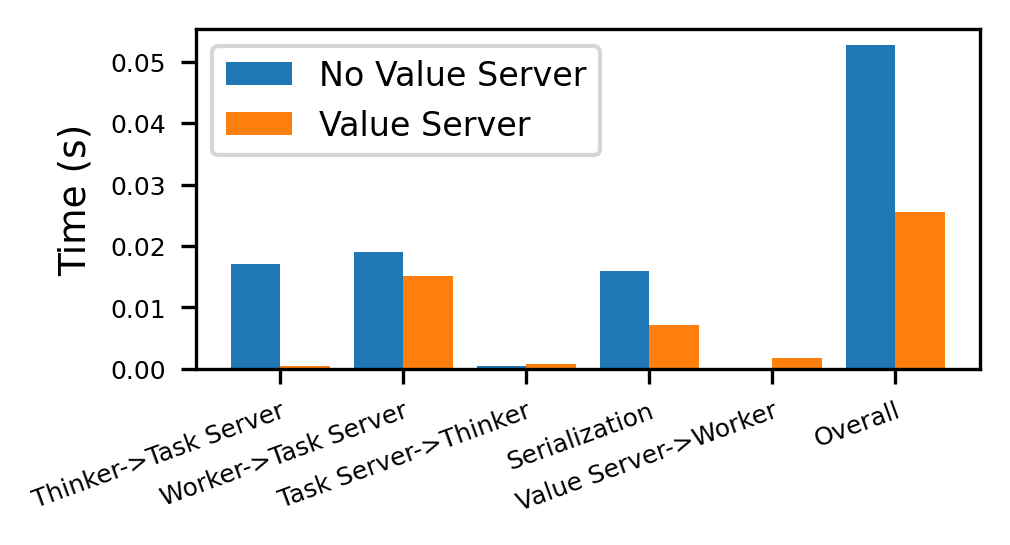}
    \caption{
        Median per-task durations for components in the Colmena task life cycle on Theta, with and without the Value Server, as measured for SynApp with eight workers, zero-length tasks, 1~MB inputs, and 0~B outputs.
        Use of the Value Server reduces time spent serializing, communicating, and deserializing task data.
    }
    \label{fig:value-server-overhead}
\end{figure}

To evaluate the impact of the Value Server,
we first run SynApp for 200 zero-length tasks with 1~MB inputs on eight nodes (i.e., \{$T$=200, $D$=0, $I$=1~MB, $O$=0, $N$=8\}), both with and without the Value Server, 
while measuring task overheads.
We see in Figure~\ref{fig:value-server-overhead} that the use of the Value Server reduces, in particular, task communication times between the Thinker client and Task Server, and serialization times.
The cost of transferring input data from the Value Server to the Worker is reduced by the use of the asynchronous data retrieval explained in \autoref{sec:architecture-communication}.
(Note that if input values were all identical, this cost would be largely eliminated due to caching.)

\begin{figure}
    \centering
    \includegraphics[width=\columnwidth,trim=2mm 2mm 2mm 2mm,clip]{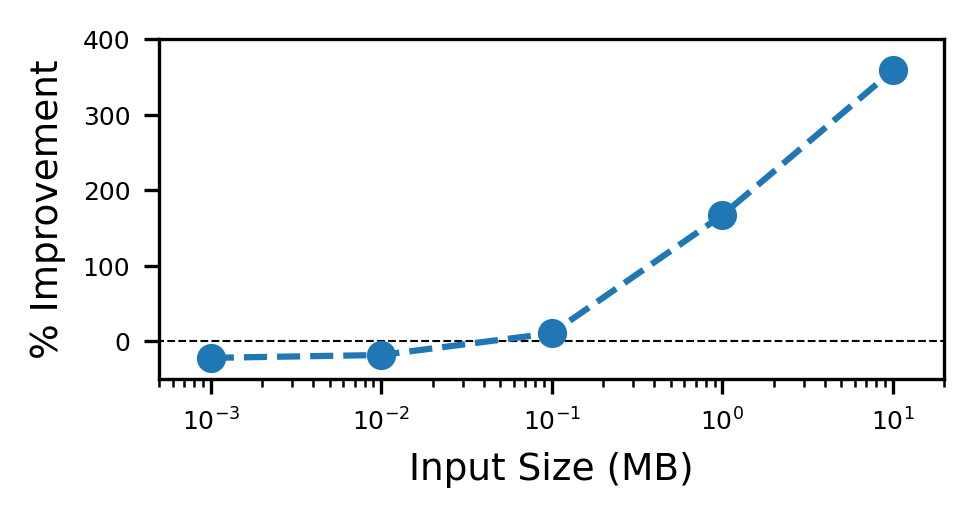}
    \caption{
        Percent reduction in SynApp overheads for configuration  
        \{$T$=200, $D$=0, $I$, $O$=0, $N$=8\} on Theta,
        with vs.\ without the Value Server, as a function of input size $I$.
        The Value Server provides performance benefits when task inputs are larger than around 0.1~MB.
    }
    \label{fig:value-server-input-size}
\end{figure}

To further study how the benefit of the Value Server varies with input size $I$, we repeat the experiments of Figure~\ref{fig:value-server-overhead} but while varying $I$ from 1~KB to 10~MB. 
The results, shown in \autoref{fig:value-server-input-size} as percentage improvement in communication overhead  time \textit{with} Value Server relative to the time \textit{without} Value Server, show that   
for small inputs ($<$10~KB), the additional cost of communicating with the Value Server is larger than the cost of passing the input data through the Task Server---but that as the input size increases, the cost of passing input data through the Task Server increases rapidly and the Value Server yields large improvements.

\begin{figure}
    \centering
    \includegraphics[width=\columnwidth,trim=2mm 2mm 2mm 0mm,clip]{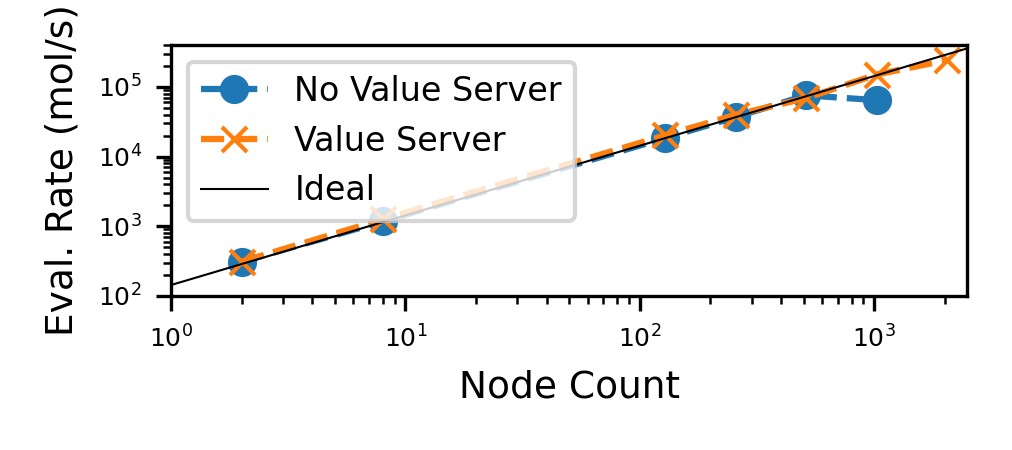}
    \caption{ML inference task performance (molecule evaluations per second) on Theta vs.\ number of nodes (one worker per node).
    The inference rate is measured starting from the time the first worker begins computation (i.e., after loading libraries) to when all inference tasks have completed.}
    \label{fig:scaling-inference-rate}
\end{figure}

\begin{figure}
    \centering
    \includegraphics[width=\columnwidth,trim=2mm 2mm 2mm 2mm,clip]{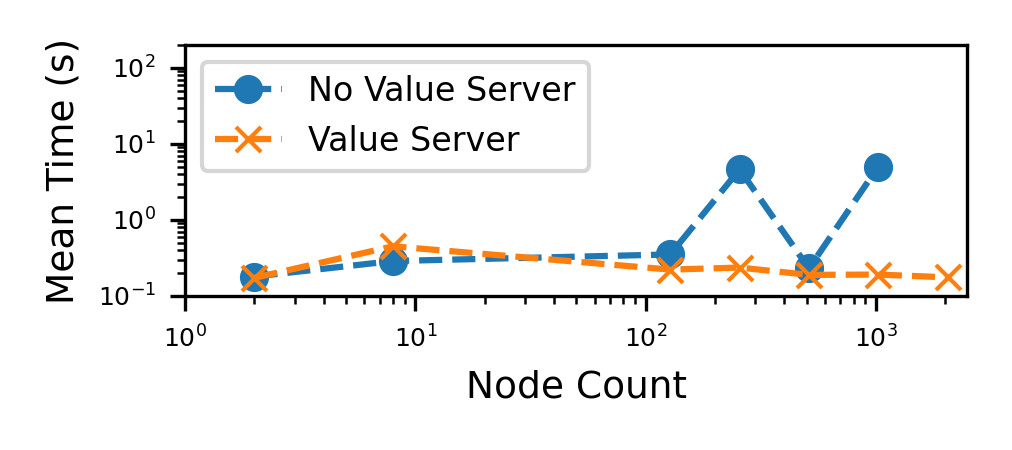}
    \caption{Time to transfer inference results from Workers to the Thinker.
        Without the Value Server, results may take up to 100~s to be sent back to the Thinker, while with the Value Server, the communication time remains more consistent.
    }
    \label{fig:scaling-transfer-time}
\end{figure}

\subsubsection{Scaling ML Assays}
As discussed in \autoref{sec:synthetic-problem}, there is a clear benefit to using the Value Server for tasks with large inputs or large results. We now study a particular sub-problem within the electrolyte design application: running machine learning inference tasks. The input task size was varied by changing the total number of molecules evaluated as a function of the number of nodes allocated for ML inference. The resulting evaluation rate, in molecules per second, is presented in Figure~\ref{fig:scaling-inference-rate}.

As noted in \autoref{sec:mdesign-scale}, individual ML inference tasks do not take long to run. Nevertheless, the Value Server can deliver significant benefits even in this case when inference results must be transferred from many nodes. 
In Figure~\ref{fig:scaling-transfer-time}, we examine the mean time taken to transfer results from Worker to Thinker, with and without the Value Server. We see that, without the Value Server, it takes up to 100~s to transfer ML inference results from more than 100 nodes. With the Value Server, however, the mean transfer time as a function of increasing node count remains constant. 

We observe significant improvements in the evaluation rate at \num{1024} nodes with the Value Server.
The time to communicate results from a completed job remains $\sim$100~ms at \num{1024} nodes when using the Value Server, in contrast to the $\sim$100~s transfer time without (Figure~\ref{fig:scaling-transfer-time}), indicating that
the workflow engine in the Task Server is not getting overloaded.
Consequently, we maintain ideal scaling up to at least \num{1024} nodes and reasonable performance at \num{2048} nodes.

\subsubsection{Performance Envelopes}

\begin{figure}
    \centering
    \includegraphics[trim=4mm 4mm 3mm 2mm,clip]{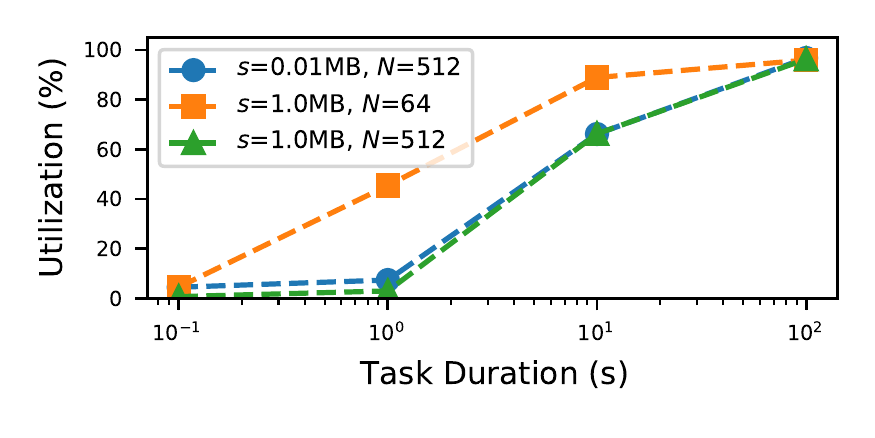}
    \caption{Performance envelope for Colmena on a Cray XC40 for different SynApp configurations. Average resource utilization changes depending on task input and output size ($I=O=s$), number of parallel workers ($N$), and average task duration.}
    \label{fig:perfenv}
\end{figure}

To explore how Colmena performance varies with task duration, result size, and number of workers, we performed additional SynApp experiments while varying those parameters. 
Figure~\ref{fig:perfenv} illustrates three performance envelopes computed on Theta. 
For example, we see that 100~s tasks that consume and produce 1~MB of data run with 90\% utilization on 512 workers.
We observe that either reducing the amount of data being transmitted or the degree of parallelism can lead to better utilization for shorter tasks.
We make this evaluation application available 
for assessing whether Colmena is an appropriate tool for different use cases or supercomputing systems~\cite{colmena}.


\section{Related Work}

Colmena requires methods for creating tasks and for monitoring and managing their execution.
It needs to support a wide range of types and scales, from multi-hour, many-node QC tasks to minute-duration, single-node ML tasks.
Accordingly, our work with Colmena builds upon much previous work. 

\textbf{General Steering Frameworks}. 
Nimrod/O~\cite{abramson2000nimrod} is an early example of a system for automated (rather than manual~\cite{mulder1999survey}) steering of simulation ensembles. 
Several general-purpose toolkits for automated steering of ensembles have been proposed in recent years, each with
different models for coupling steering and simulation.
Here, we describe their key features and how they inspired our development of Colmena.

The DeepHyper~\cite{balaprakash2018deephyper} hyperparameter optimization system uses a centralized planning process to select tasks and distributes a single type of assay across multiple nodes using a workflow engine. The planning process submits tasks to the workflow engine (DeepHyper supports Balsam~\cite{salim2019balsam} and Ray~\cite{mortiz2018ray}) and queries the engine for completed results.
Supervisor~\cite{wozniak2018supervisor} has a similar centralized architecture to DeepHyper, with a single node for communicating with the high-performance Swift/T~\cite{wozniak13swiftt} workflow engine via a queue.
Proxima~\cite{zamora2021proxima} uses ML methods to dynamically tune a surrogate-modeling configuration in response to real-time feedback from the ongoing simulation, but does not optimize for use of HPC.
Colmena uses a similar centralized model for task planning; is designed, like DeepHyper, to support multiple workflow engines; and provides easier support for multiple task types.

LibEnsemble~\cite{libEnsemble} expresses ensembles using a model where tasks produced by \textit{task generation} (akin to steering) workers are executed by \textit{simulation} workers, and simulation results are fed back to inform the generation tasks. The steering policy can be either centralized (single task generator) or decentralized (multiple task generators). A manager service deploys the task generation and simulation workers onto multiple nodes and routes data across workers. Colmena shares LibEnsemble's ability to dedicate more than one node to steering-related tasks and provides the ability to break policies into an unlimited number of worker types.

Ray~\cite{mortiz2018ray} uses a similar distributed model to LibEnsemble but permits ensembles to use many types of interacting agents (beyond generator and simulator) with complex coordination policies. Colmena has a similar agent-oriented programming model but centralizes all agents to a single node so that they can communicate with shared memory.

Rocketsled~\cite{dunn2019rocketsled} expresses the steering task as a step within a Fireworks~\cite{jain2015fireworks} workflow that can add new nodes to the workflow graph. Colmena uses a different programming model where planning tasks need not be triggered by workflow events and was designed to separate planning logic from the task execution engine. 
Cray SmartSim~\cite{partee202smartsim} allows multiple planning scripts to independently launch jobs on a cluster and coordinate among each other using Redis. The Colmena and SmartSim programming models are similar, in that multiple planning scripts can submit and receive work concurrently, and possible tasks can be enumerated as a list of assays. Colmena has utilities that simplify building planning strategies for ensemble steering (e.g., pulling from result queues, event-triggered resource reallocation), whereas SmartSim is capable of building other types of AI+simulation applications (e.g., online analysis, inference from simulation codes).

In short, the Colmena toolkit is purpose-built for expressing the complex policies needed to make efficient use of highly parallel supercomputers for computational campaigns.
We designed Colmena to provide many of the features of current frameworks for steering computational campaigns, including the centralized programming model of codes such as DeepHyper or Supervisor, the ability to deploy planning tasks across multiple nodes illustrated by libEnsemble, and simple routes to building planning policies as cooperative agents in the style of Ray and SmartSim.
Colmena presents a single package able to recreate the parallelization strategies of all current steering frameworks and provides the flexibility needed to explore even more sophisticated approaches.

\textbf{Active Learning on HPC}.
Active learning methods obtain 
new training data via online querying of an information source, including (as here) computational simulations. The approaches just reviewed may be viewed as active learning methods.


\textbf{Reinforcement Learning on HPC}. Algorithms for training reinforcement learning models have much in common with those for steering ensemble simulations.
Reinforcement learning agents gather data by using a policy to guide the evolution of different environments (e.g., simulations for physical systems) and periodically retrain this policy to better steer the environments towards desirable states.
Computations that simulate environments, make policy decisions, and retrain policies can all occur asynchronously across distributed resources, akin to Colmena's Thinker/Task Server model for ensemble simulations.
Consequently, the design of steering algorithms is related to approaches for distributed training of reinforcement learning (e.g., IMPALA~\cite{espeholt2018impala}) and to toolkits for deploying reinforcement learning at scale (e.g., RLLib~\cite{rllib}, ExaRL~\cite{exarl}).
Such algorithms for training reinforcement learning policies are a class of approaches that can be expressed with Colmena.

\textbf{Specialized Steering Frameworks}. Specialized applications that dynamically create or reorder tasks are also prevalent in the literature.
Various domain-specific tools (e.g., XtalOpt~\cite{lonie2011xtalopt}, Kombine~\cite{kombine}, DeepDriveMD~\cite{lee2019deepdrivemd}, CAMD~\cite{montoya2020camd}) engage different patterns for generating tasks in solving different classes of problems (e.g., optimization, parameter estimation).
Tools that perform hyperparameter searches for neural network design, such as DeepHyper~\cite{balaprakash2018deephyper} and Tune~\cite{liaw2018tune}, illustrate how to handle ML tasks at scale.
Colmena adapts concepts from these tools; each of the algorithms in these tools can be implemented using Colmena.

\textbf{Workflow Systems}. Colmena also relies heavily on workflow systems to distribute computations across many nodes. 
Applicable workflow systems include Ray~\cite{mortiz2018ray}, Balsam~\cite{salim2019balsam}, RADICAL Cyber Toolkit~\cite{radical2019}, Parsl~\cite{babuji19parsl}, Fireworks~\cite{jain2015fireworks}, and many others~\cite{workflowlist}.
Such systems provide unique approaches to specifying tasks and runtime systems for executing them across distributed resources.
Colmena is designed to make use of workflow tools and not to make any new contributions in the design of workflow systems.

\textbf{Process Management Systems}. Methods for deploying many concurrent, short-duration tasks is another area of active research.
The challenges are well explained in a recent work that studied many-task performance on Summit~\cite{turilli2019characterizing}.
The Process Management Interface (PMIx)~\cite{castain2018pmix} defines an API for such capabilities that is available on some supercomputers (e.g., Summit).
There are also efforts, such as MPI\_Comm\_launch~\cite{wozniak2019mpi}, to incorporate process management methods into the Message Passing Interface.
Systems for launching, monitoring, and managing large numbers of tasks on HPC are going to be critical as we scale Colmena to larger computational resources.

\section{Conclusions}

We introduced Colmena, an open-source Python library for machine-learning-based steering of ensemble computations on HPC systems.
We first formalized the steering process as a design problem where one must decide which computations to 
perform on what inputs to produce a record of simulations with maximal value at minimal cost.
We then described how Colmena facilitates building such steering applications by permitting the construction and composition of a Thinker that implements the decision making processes used to define computational  
tasks and a Task Server that distributes execution across HPC resources.
We illustrated the use of the Colmena library with a molecular design application that finds molecules with high resistance 
to oxidation at rates 100$\times$ faster than naive searches by interleaving simulation and ML tasks.
We demonstrated effective scaling on up to \num{1024} nodes (\num{65536} cores) and illustrate how to improve the scaling of the application further by using a separate subsystem for transferring large result objects.

We intend that Colmena provides a toolkit for exploring methods for steering ensemble simulations.
The flexible, multi-threaded Thinker class permits implementing varied, complex policies for interleaving different types of computation.
Our primary goal for creating Colmena is to support the expression of steering policies that use ML to augment human intelligence in designing and managing computational campaigns.
Interfaced with a Task Server built using Parsl, users can execute these policies at large scales
and across heterogeneous computing resources.
As we learn more, we will build templates for common classes of decision problems (e.g., model-based optimization, reinforcement learning)
that allow users to quickly deploy state-of-the-art steering policies.
Through this work, we hope to enable computational campaigns that take fuller advantage of current and next-generation supercomputers.



\section*{Data and Software Availability}

The source code used in the this manuscript, logging information for each of the runs described in the paper, and Jupyter notebooks used to analyze logs and produce figures are all published via the Materials Data Facility.\cite{colmena_data,blaiszik2019mdf}
The source code and Jupyter notebooks associated with this manuscript is also available on GitHub at \url{https://github.com/exalearn/electrolyte-design/}, which will be updated as our work proceeds.

\section*{Acknolwedgements}

LW, GS, GP, RC, RT, and IF acknowledge support by the ExaLearn Co-design Center of Exascale Computing Project (17-SC-20-SC), a collaborative effort of the U.S. Department of Energy Office of Science and the National Nuclear Security Administration, to develop Colmena and evaluate its performance on HPC.
YB and KC were supported to integrate Parsl support into Colmena by NSF Grant 1550588 and ExaWorks Project within the Exascale Computing Project.
GP and KC were supported to develop the value server by NSF Grant 2004894.
LW, ND, PCR, RSA, and LAC were supported to define the electrolyte design problem and develop the computational workflows need to solve it by the Joint Center for Energy Storage Research (JCESR), an Energy
Innovation Hub funded by the US Department of Energy, Office of Science, Basic Energy Sciences.
This research used resources of the Argonne Leadership Computing Facility (ALCF), which is a DOE Office of Science User Facility supported under Contract DE-AC02-06CH11357, and was supported by the ALCF Data Science Program.

\bibliographystyle{ieeetran}
\bibliography{base.bib,MAIN.bib}

\end{document}

%% file: authors.tex
\author{
    \IEEEauthorblockN{Logan Ward,\IEEEauthorrefmark{1}\IEEEauthorrefmark{3} Ganesh Sivaraman,\IEEEauthorrefmark{1}
    J. Gregory Pauloski,\IEEEauthorrefmark{2} Yadu Babuji,\IEEEauthorrefmark{2} Ryan Chard,\IEEEauthorrefmark{1} \\
    Naveen Dandu,\IEEEauthorrefmark{3} Paul C. Redfern,\IEEEauthorrefmark{3} Rajeev S. Assary,\IEEEauthorrefmark{3}
    Kyle Chard,\IEEEauthorrefmark{2} Larry A. Curtiss,\IEEEauthorrefmark{3} \\
    Rajeev Thakur,\IEEEauthorrefmark{1} and Ian Foster\IEEEauthorrefmark{1}\IEEEauthorrefmark{2}}
    \IEEEauthorblockA{\IEEEauthorrefmark{1}Data Science and Learning Division, Argonne National Laboratory, Lemont, IL, USA}
    \IEEEauthorblockA{\IEEEauthorrefmark{2}Department of Computer Science, University of Chicago, Chicago, IL, USA}
    \IEEEauthorblockA{\IEEEauthorrefmark{3}Joint Center for Energy Storage Research, University of Chicago, Chicago, IL, USA}
}